\documentclass[twocolumn,secnumarabic,amssymb, nobibnotes, aps, prd]{revtex4-1}

\usepackage{graphicx,array,multirow}

\begin{document}

\title{Search for nucleon-nucleon correlations in neutrino-argon scattering}%

\author{Kajetan Niewczas and Jan T. Sobczyk}%
\email{jsobczyk@ift.uni.wroc.pl}
\affiliation{Institute of Theoretical Physics, University of Wroc\l aw, pl. M. Borna 9,
50-204, Wroc\l aw, Poland}
\date{\today}%
\begin{abstract}
A sample of two proton and no pion events selected in the ArgoNeuT neutrino scattering experiment on a liquid argon target 
[Phys. Rev. D90 (2014) 012008]
is analyzed with the NuWro Monte Carlo event generator. An attempt is made to estimate how likely it is to obtain observed numbers 
of laboratory frame and reconstructed \mbox{back-to-back} nucleon pairs. For laboratory frame \mbox{back-to-back} events a clear data/MC
discrepancy is seen. For the reconstructed nucleon pairs a good agreement is reported. We provide a simple kinematical argument
why this accordance is expected.
\end{abstract}

\maketitle

\section{Introduction}
\label{sec:intro}

There is a lot of discussion about systematic errors in the next round of neutrino oscillation experiments, such as 
DUNE~\cite{Adams:2013qkq} and HyperKamiokande~\cite{Abe:2015zbg}. The measurement of CP violation in the leptonic sector of the Standard Model 
is a very challenging goal and sensitivity studies suggest that it is desirable to reduce systematic errors in this measurement 
to the level of $1-3\%$. One of the most important sources of systematical error is the neutrino-nucleus interaction cross sections. 
In the $1-5$~GeV energy region they are known with a precision not exceeding $20\%$. In actual experimental setup thanks to information 
from a near detector many uncertainties cancel but significant experimental and theoretical effort to understand better neutrino interactions 
is necessary.

The most important dynamical mechanism in the discussed energy region is charge current quasielastic scattering (CCQE):

\begin{eqnarray}
\label{eq:ccqe}
\nu_l\ n\rightarrow l^-\ p,\ \ \ \ \ \bar\nu_l\ p\rightarrow l^+\ n
\end{eqnarray}
where $l\in\{ e,\mu ,\tau\}$, and $n$ and $p$ denote the neutron and proton. 
In a case of neutrino-nucleus reaction in the impulse appriximation regime, 
it is difficult to separate CCQE from other processes giving rise to similar final states. The main background processes 
are pion production 
with its subsequent absorption and two-body current scattering with virtual boson absorbed on a correlated nucleon-nucleon pair. 
Pion absorption is an example of final state interactions (FSI): secondary interactions of hadrons inside the nucleus. 
FSI and other nuclear effects like Fermi motion and binding energy make the measurement of axial form-factors entering the 
nucleon-nucleon 
weak current matrix element difficult.

The two-body current contribution to the neutrino inclusive cross section is a subject of many theoretical studies following MiniBooNE measurement 
of the large effective axial mass~\cite{AguilarArevalo:2010zc}, the main unknown in axial form factor. 
Since the pioneer work by Martini et al~\cite{Martini:2009uj} several models were proposed to explain the
MiniBooNE result~\cite{Nieves:2011pp, Amaro:2011qb}. All of them provide predictions for a multinucleon ejection contribution 
to $\nu_\mu$ inclusive cross section. As such, they can be confronted with results for both neutrino and antineutrino scattering 
in MiniBooNE and also more recent MINERvA experiments~\cite{Fiorentini:2013ezn, Fields:2013zhk}. The picture that arises is not completely clear. 
Further studies are needed and in particular it seems necessary to look also at final state nucleons resulting 
from two-body current processes~\cite{Sobczyk:2012ms}. 
Promising ab initio nuclear physics computations of two body current can provide nuclear response functions for light nuclei though 
in a limited phase space of $q<800$~MeV ($q$ is three-momentum transfer)~\cite{Lovato:2014eva}.

Good understanding of neutrino scattering in the quasielastic peak region is a prerequisite in attempts to reduce cross section 
systematic error in oscillation experiments. It is very important to have experimental data allowing for better control 
of two-body current contribution and FSI effects. It is particulary important to have such a data for argon target, 
because liquid argon is going to be used in the short baseline experiment at Fermilab~\cite{Antonello:2015lea} and also in the DUNE experiment. 

Recently, interesting argon target experimental results were reported by ArgoNeuT Collaboration \cite{Acciarri:2014gev}. 
The liquid argon detector technology allows for reconstruction of proton tracks with momenta as low as $200$~MeV/c, 
i.e. below argon Fermi momentum. It becomes possible to investigate nontrivial nuclear effects: nucleon rescatterings, 
pion absorption mechanism and also two-body current interaction. 

The ArgoNeuT data are of low statistics and it is difficult to draw out of them definite conclusions. 
However, much higher statistics data will soon become available from the MicroBooNE experiment. 
It seems important to perform Monte Carlo studies inspired by the ArgoNeuT results in order to identify promising observables 
in future experiments. ArgoNeuT investigated a sample of events with no pions and two reconstructed protons. 
ArgoNeuT main interest is a search for the short-range correlated (SRC) nucleon pairs within the argon nuclei. 
Two interesting observables have been identified. The first one is a distribution of angles between two protons in the laboratory frame. 
It was noticed that there are many proton pairs in almost \mbox{back-to-back} configuration. ArgoNeuT also looked for two body current events 
trying to reconstruct initial nucleon-nucleon configuration. The conclusion was that there is an excess of events 
with \mbox{back-to-back} initial nucleon-nucleon state. The goal of this paper is to analyze ArgoNeuT discoveries 
using NuWro Monte Carlo event generator \cite{Golan:2012wx}. The main question is whether physical models implemented in NuWro are
sufficient to explain the ArgoNeuT results. Or, perhaps the ArgoNeuT discoveries point to more sophisticated nuclear physics 
that should be included in neutrino MC simulation tools.

The paper is organized as follows: in Section~\ref{sec:argoneut}, we provide basis information about ArgoNeuT experiment; 
then in Section~\ref{sec:nuwro}, we describe NuWro MC generator reviewing physical models that are implemented there;
Section~\ref{sec:hammer} is devoted to {\it hammer} events with two \mbox{back-to-back} protons in the laboratory frame;
in Section~\ref{sec:recon}, we follow ArgoNeuT prescriptions to identify initial two nucleon configuration in events 
that were likely to occur on correlated nucleon-nucleon pairs; and in the final two Sections,~\ref{sec:discussion} and 
~\ref{sec:conclusions}, we discuss the results and present our conclusions, respectively.

\section{ArgoNeuT experiment}
\label{sec:argoneut}

ArgoNeuT features a Liquid Argon Time Projection Chamber (LArTPC) immersed in a 550 liter vacuum-insulated cryostat. 
The detector has an active volume of $47$x$40$x$90$~cm$^3$ (170 liters); a rectangular box filled up with 240 kg of LAr. 
The neutrino beam is led along the longest dimension of the chamber. ArgoNeuT operated during two runs of different horn configurations 
of the NuMI LE (low energy option) beam from September 2009 to February 2010 at the Fermi National Accelerator Laboratory (FNAL). 
It is the only LArTPC to operate on low energetic beam (0.1 -- 10~GeV). The first run, ~2 weeks long [$8.5\cdot10^{18}$ 
proton on target (POT)], in the $\nu$-beam mode acquired 729 neutrino charge current events. 
The second run, ~5 month long [$1.25\cdot10^{20}$ POT], in the $\bar{\nu}$-beam mode with a large \textit{neutrino} 
fraction acquired 3759 neutrino charge current events. The average neutrino energy in the first beam was 
$\langle E_\nu \rangle \simeq 4$~GeV, while in the second beam it was $\langle E_\nu \rangle \simeq 10$~GeV. The overall number of
729+3759 neutrino events in both modes is efficiency corrected.

During the experiment the detector was set slightly off beam axis (TPC center located 26cm below the beam plane). The 
beam direction is the $\hat{z}$ axis in the laboratory frame of reference.

Charged particles crossing the active volume ionize free electron tracks that 
drift under the uniform electric field ($481$~V/cm) along the horizontal $\hat{x}$ direction. The electron track image is obtained 
by collecting signals from two wire-planes situated on the right TPC wall. The maximal drift length is $47$~cm. 
The wire-planes, both made from 240 wires each and rotated $\pm 60^\circ$ to the beam plane, allow 
for the identification of signals from individual wires. 
As in liquid argon there is no charge multiplication, the signal pulse height is proportional to the amount of ionization charge 
in the track segment. Therefore summing the charge over the entire track length gives us the calorimetric information. 
Combining data from both wire-planes, that have drift coordinate in common, one can fully reconstruct three-dimensional image of the event. 
For particles having contained tracks within the TPC energy loss is a function of distance. This dependence is a powerful tool 
for the particle identification. For uncontained muons escaping in the forward direction, the momentum and charge identification 
was performed using the MINOS Near Detector (MINOS-ND) calorimeter located downstream from ArgoNeuT.

The technology used allows us to obtain a very low proton kinetic energy detection threshold of $T^{thr}_p = 21$~MeV, or $200$~MeV/c 
of momentum. 

\section{NuWro simulations}
\label{sec:nuwro}

NuWro is a versatile Monte Carlo neutrino event generator developed over the last 10 years at the Wroclaw University. 
It provides a complete description of \mbox{(anti-)}neutrino interactions on arbitrary nucleon/nucleus targets in the energy range 
from $\sim 100$~MeV to $\sim 1$~TeV. Basic neutrino interaction modes on free nucleon target are: 
\begin{itemize}
 \item CCQE (including the elastic analog of Eq. (\ref{eq:ccqe}) for neutral current reaction),
 \item RES (from resonant), covering a region of invariant hadronic mass $W\leq 1.6$~GeV; the dominant RES process is  
 $\Delta$ resonance excitation
\begin{eqnarray}
\label{eq:delta}
\nu_l\ N\rightarrow l^- \Delta
\end{eqnarray}
with $N$ standing for either proton or neutron,
 \item DIS (a slightly misleading neutrino Monte Carlo community jargon): all the inelastic processes with $W\geq 1.6$~GeV.
\end{itemize}
In the case of neutrino-nucleus scattering two other interaction modes are
\begin{itemize}
 \item COH -- coherent pion production,
 \item MEC -- two-body current processes (MEC stands for meson exchange current; some authors call this mechanism 
 multinucleon ejection or np-nh, n-particles and n-holes in the language of many body theory).
\end{itemize}

Neutrino-nucleus CCQE, RES, DIS and MEC reactions are modeled as a two-step proces; the primary interaction on one or two nucleons 
is followed by final state interactions. NuWro FSI effects are described by custom made 
semiclassical intranuclear cascade (INC) model~\cite{Golan:2012wx}. It includes pion absorption treated according to the model
of Oset et al \cite{Salcedo:1987md}.

NuWro is equipped with a detector interface and can be used in experimental studies. For the purpose of this analysis 
the only relevant detector effect that must be taken into consideration is the detector finite size. A requirement is that 
proton tracks are fully contained in the detector, and this eliminates a fraction of proton long track events. 
In order to simulate the detector finite size effects, we use the ArgoNeuT algorithm to calculate particle kinetic energy 
$T(R)$ (in the units of MeV) based on its track length $R$ (in the units of cm)~\cite{Acciarri:2013met}:

\begin{eqnarray}
\label{eq:tracklength}
T(R) = \frac{A}{b+1} R^{b+1},
\end{eqnarray}
where the parameters for proton are $A = 17$ in the units of MeV/cm$^{(1+b)}$ and $b = -0.42$. 
For the reader's orientation, protons with momentum $500$~MeV/c 
travel average distance of $12.2$~cm. For each event we uniformly draw a position of occurance within the TPC ($47$x$40$x$90$~cm$^3$). 
Then using Eq. (\ref{eq:tracklength}) we calculate the length of track for each proton. Attaching the length to the actual 
proton momentum directions, we can decide if the track is fully contained in the TPC. Events with uncontained proton tracks are discarded.

The fluxes (NuMI LE) used in the simulations were provided to us by the ArgoNeuT Collaboration. In the analysis ArgoNeuT Collaboration 
used neutrino events from both (neutrino and antineutrino) runs. NuWro simulation normalization is defined not by numbers 
of two proton events in both modes identified by ArgoNeuT but rather by the overall numbers of neutrino charge current events in two runs. 
The numbers of such events are 729 and 3759 in $\nu_\mu$ and $\bar\nu_\mu$ modes respectively, so that in the NuWro simulation 
we produced 2916000 and 15036000 neutrino charge current events in two modes to keep their relative fractions fixed.

\subsection{NuWro configurations}

NuWro offers a lot of flexibility for the composition of models used in an actual study. In this analysis we use two NuWro configurations.

The first one is the default NuWro configuration: 

\begin{itemize}
 \item CCQE
 \begin{itemize}
  \item local Fermi gas,
  \item BBBA vector form factors,
  \item dipole axial form factor with $M_A=1.03$~MeV,
  \item no coherent length effects for outgoing nucleon.
 \end{itemize}
 \item RES
 \begin{itemize}
  \item N-$\Delta$ axial form-factor in dipole parameterization with $M_A=0.94$~GeV, $C^5_A(0)=1.19$~\cite{Graczyk:2009qm},
  \item nuclear target pion production reduced due to $\Delta$ in-medium self-energy implemented in the approximate way 
  using results of~\cite{Sobczyk:2012zj},
  \item non-resonant background added incoherently~\cite{Juszczak:2005zs},
  \item $\Delta$ finite life-time effects~\cite{Golan:2012wx},
  \item angular distribution of pions resulting from $\Delta$ decays modeled using results of ANL and BNL 
  experimental measurements~\cite{Sobczyk:2014xza},
 \end{itemize}
 \item DIS 
 \begin{itemize}
  \item PYTHIA fragmentation routines,
  \item formation zone effects modeled as explained in~\cite{Golan:2012wx},
 \end{itemize}
 \item MEC 
 \begin{itemize}
  \item Nieves et al model with a momentum transfer cut $q\leq 1.2$~GeV/c,
  \item in $95\%$ of events interaction occurs on correlated \mbox{back-to-back} proton-neutron pairs,
  \item finite state nucleons are assigned momenta using the phase space model~\cite{Sobczyk:2012ms},
  \item no coherence length/formation zone effects for outgoing nucleons.
 \end{itemize}
\end{itemize}

In the second configuration, the LFG model for CCQE is replaced by the hole spectral function (SF) approach~\cite{Benhar:1989aw}. 
In the SF approach one introduces a realistic distribution of target nucleon momenta and excitation energies. 
The target nucleon momentum distribution contains a high momentum tail coming from correlated nucleon-nucleon pairs. 
In the NuWro implementation of the SF approach one distinguishes if an interaction occurs on a nucleon discribed 
by a mean field approach or on a nucleon forming a correlated pair. In the second case, it is assumed that there is also 
a correlated nucleon that does not participate in the interaction but, after initial interaction, propagates inside nucleus. 
Its initial momentum is assumed to be opposite (as a three-vector) to that of the interacting nucleon. 

We are aware that from the theorist perspective it is not fully consistent to combine the SF approach and 
the MEC model of Nieves et al~\cite{Benhar:2015ula}. On the other hand, 
both dynamical mechanisms provide events originating from correlated nucleon pairs, 
and it is important to investigate the effect
seen after reconstruction procedures proposed by the ArgoNeuT Collaboration are applied.

NuWro MC generator shares many common features with Monte Carlo generators 
NEUT and GENIE used by experimental groups (for a review on neutrino generators see~\cite{PDG-MC}). 
Therefore, conclusions about NuWro performance with respect to
the ArgNeuT data are likely to be applicable also to other MCs.

\subsection{Two proton events}

ArgoNeuT performed a detail analysis of events with no pions and exactly two protons detected in the final state. 
Strictly speaking, the investigated data sample contains no charged pions with kinetic energy lower than $10$~MeV, but according
to NuWro the difference between two selections is negligible.
To follow ArgoNeuT considerations, we define NuWro samples of events in the same way. We allow for an arbitrary number 
of undetectable knocked out neutrons. ArgoNeuT collected 30 events of this type. The details about our analysis 
can be found in table~\ref{tab:statistics}. Our choice for the 
relative numbers of events from $\nu$ and $\bar\nu$ modes was explained before. The $3.4\%$ fraction of two-proton events 
in the ArgoNeut data sample was estimated using efficiency corrections \cite{Acciarri:2014gev}.

Using NuWro, we checked that the distribution of muon angles for two-proton events is strongly peaked at $\sim 5^o$ without much
difference for separated dynamical mechanisms. The contribution from muons with angles larger than $30^o$ is very small, of the 
order of 3.5\%. We did not introduce any muon efficiency corrections because for angles $<30^o$ ArgoNeuT muon detection efficiency is 
approximately constant and equal $\sim 90\%$.

\begin{table}[!htb]
\resizebox{.5\textwidth}{!}{
\begin{tabular}{l|cc|cc|c|}
\cline{2-6}
\rule{0pt}{2.2ex} & \multicolumn{2}{c|}{$\nu$-mode} & \multicolumn{2}{c|}{$\bar{\nu}$-mode} & \% of all 2p events \\
\rule{0pt}{2.2ex}  & \multicolumn{4}{c|}{(\% of investigated 2p events)} & in total sample \\ \hline
\multicolumn{1}{|l|}{ArgoNeuT} & \rule{0pt}{2.2ex} 11 & (37\%) & 19 & (63\%) & 3.4\% \\ \hline
\multicolumn{1}{|l|}{NuWro: LFG} & \rule{0pt}{2.2ex} 57979 & (21.9\%) & 206955 & (78.1\%) & 4.7\% \\ \hline
\multicolumn{1}{|l|}{NuWro: SF} & \rule{0pt}{2.2ex} 61910 & (22.1\%) & 217982 & (77.9\%) & 4.9\% \\ \hline
\end{tabular}
}
\caption{Two-proton sample statistics from both ArgoNeuT and NuWro. The last column shows the fraction 
of the two-proton events without detector effects. 
The previous two columns show the contributions of two-proton events from both neutrino beams to the investigated subsample with detector effects.}
\label{tab:statistics}
\end{table}

NuWro predictions are in reasonable agreement with the ArgoNeuT results. 
Different relative contributions from $\nu$ and $\bar\nu$ modes can be merely a statistical fluctuation.

In the next sections we are comparing to the ArgoNeut data for two proton events that is not efficiency corrected, and
in the NuWro results we consider only events with the muon angle $<30^o$.

\section{Hammer events in the laboratory frame}
\label{sec:hammer}

The first interesting ArgoNeuT observable is a distribution of the cosine of the angle $\gamma$ between two proton three-momenta
in the laboratory frame. 
ArgoNeuT found an intriguing enhancement in the number of, so called, hammer events, that are proton pairs 
in almost \mbox{back-to-back} configuration in the final state, defined as $\cos(\gamma) \leq -0.95$.

\begin{figure}[!htb]
\centering\includegraphics[width=\columnwidth]{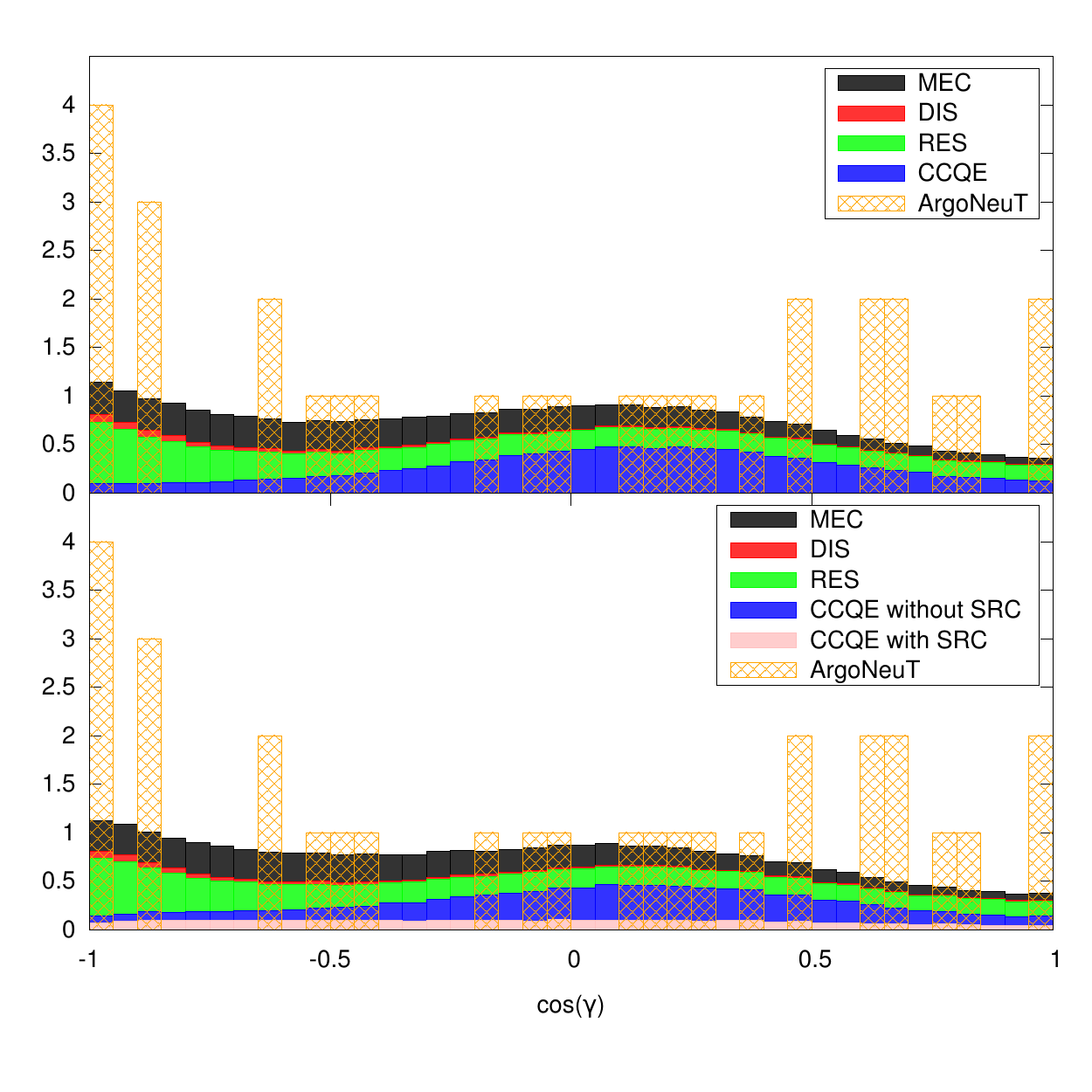}
\caption{(Color online) Distribution of the cosine of the angle between two protons in the final state. (Top) LFG model, 
(bottom) SP approach.}
\label{fig:cos_g}
\end{figure}

In NuWro we calculated the distribution of $\cos\gamma$ in two-proton events and in Fig.~\ref{fig:cos_g} 
we compared them with the experimental data. The NuWro results are normalized to the same area.

NuWro distributions are rather flat with two not very pronounced maxima: the first one at $\cos \gamma\sim -1$ 
and the second one at  $\cos \gamma\sim 0$
According to NuWro, most of the hammer events come from the RES and MEC mechanisms.

Using the NuWro distributions we can calculate the probability of obtaining 4 or more events $P(4+)$ in the first bin. 
We treat the NuWro results as the probability distribution and use Poisson statistics. Our results are:
\begin{itemize}
\item $P(4+)=2.9\%$ for the LFG model,
\item $P(4+)=2.6\%$ for the SF approach.
\end{itemize}
The probabilities are similar and in both cases rather small. From the NuWro perspective the appearance of as many as four hammer events 
in a sample of 30 two protons events is an interesting fact. The probability that it is merely a statistical fluctuation 
is only about 3\%. Certainly, better statistics data is required in order to draw a definite conclusion 
that MC event generators are unable to understand appearance of so many hammer events.

ArgoNeuT proposed also to study a subsample of two proton events by demanding that both protons 
have momenta larger than argon Fermi momentum. In this way, they received a reduced sample of only 19 events. 
In NuWro simulations, the similar requirement reduces the number of events by a factor of 30\%, in very good agreement with 
11/30 in the ArgoNeuT study.

\begin{figure}[!htb]
\centering\includegraphics[width=\columnwidth]{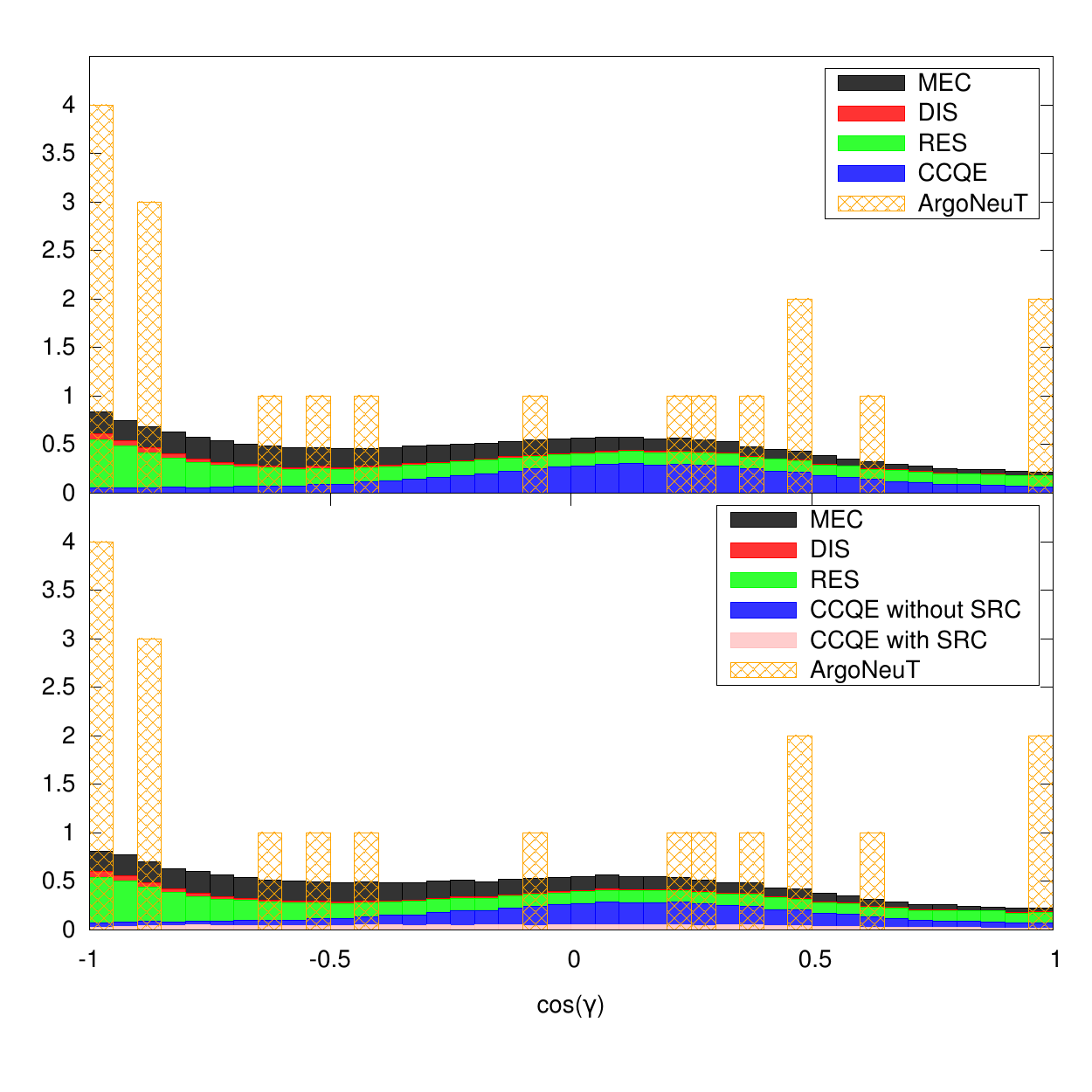}
\caption{(Color online) Distribution of the cosine of the angle between two protons in the final state. 
Subsample with both protons momenta above argon Fermi momentum. (Top) LFG model, (bottom) SP approach.}
\label{fig:cos_g_kf}
\end{figure}

In Fig.~\ref{fig:cos_g_kf}, we show the comparison of experimental and NuWro results with the additional constraint on the proton momenta. 
We see that the shape of the NuWro distribution did not change significantly.

We calculated again the probability of having four or more events $P(4+)$ out of 19 in the first bin using 
NuWro results as the probability distribution. Our results are
\begin{itemize}
\item $P(4+)=1.1\%$ for the LFG model,
\item $P(4+)=0.9\%$ for the SF approach.
\end{itemize}
Both probabilities are lower than before. Additionally, the detected hammer events were found to have both protons with similar momenta, 
i.e. $|\vec{p}_1| \simeq |\vec{p}_2| > k_F$ (by definition $p_1$ is more energetic than $p_2$: ${|\vec{p}_1|} \geq {|\vec{p}_2|}$). 
Moreover, the events were characterized by typical values of missing transverse momentum $|p^T_{miss}| \geq 300$~MeV/c, where $p^T_{miss}$ 
is defined as the length of the sum of three-momenta of all detectable particles
(muon, protons) in the plane perpendicular to the beam direction. ArgoNeuT gives the explanation that laboratory frame hammer events originate mostly from the RES mechanism~\cite{Bellotti:1973vn}. This agrees with the breakdown of NuWro events in various interaction modes. RES events contributing to two proton final states are those with pion being produced and subsequently absorbed inside nucleus. While NuWro agrees with the ArgoNeuT on the dominant mechanism leading to hammer events, it cannot explain the fact that so many hammer events are contained in the samples of 30 or 19 ArgoNeuT events. 

\section{Reconstructed \mbox{back-to-back} nucleons before interaction}
\label{sec:recon}

The ArgoNeuT Collaboration tried to identify a subsample of events occuring on correlated 
nucleon-nucleon pairs. ArgoNeuT proposed a procedure to reconstruct nucleon initial state 
configuration before the interaction assuming that it was a two nucleon state.
The sample of 19 events discussed in~\ref{sec:hammer} is further reduced by subtracting four (most likely RES) hammer events. 

The incident neutrino energy and four-momentum transfered to the hadronic system are not known on event-by-event basis. 
The precise reconstruction of their values is not possible because of the FSI effects blurring the image. 
ArgoNeuT attempted to approximate the nucleus recoil energy 
with the formula $T_{A-2} \approx \frac{(\vec{p}^T_{miss})^2}{2 M_{A-2}}$ ($M_{A-2}$ is large enough and non-relativistic formula is a 
good approximation). Then the neutrino energy was reconstructed as
\begin{eqnarray}
\label{eq:reconstruction}
E_\nu = E_\mu + T_{p1} + T_{p2} + T_{A-2} + E_{miss},
\end{eqnarray}
where $T_{p1}$ and $T_{p2}$ are proton kinetic energies and $E_{miss} = 30$~MeV is 
the approximate energy needed to knock out a nucleon pair from an argon nucleus. 
With the reconstructed neutrino energy and information about the final state muon one can calculate three-momentum transfer. 
The final ansatz is that the three-momentum transfer was absorbed by the most energetic final state proton only and both
protons did not suffer from the FSI effects. In this way one gets the initial state nucleon three-momenta 
and in particular the angle $\gamma^i$ between both nucleons in the initial state. 
For events not occuring on nucleon-nucleon pairs the $\gamma^i$ reconstruction procedure has no physical meaning.

ArgoNeuT found three reconstructed nucleon-nucleon pairs in approximately \mbox{back-to-back} configuration defined as $\cos\gamma^i\leq -0.9$. 
In the Fig.~6 in~\cite{Acciarri:2014gev} three of them are shown in the bin $(-0.95, -0.9)$. 
ArgoNeuT also discussed the fourth event which lies on the bin boundary with $\cos\gamma^i\sim -0.89$. 
There are altogether six events in the region of $\cos\gamma^i\leq -0.8$.

Following the ArgoNeuT procedures we took the NuWro sample of two proton events and subtracted 

\begin{itemize}
 \item hammer events in the LAB frame,
 \item events with less energetic proton momentum smaller than argon Fermi momentum.
\end{itemize}

For the remaining NuWro events we performed the ArgoNeuT reconstruction as described above. 

\begin{figure}[!htb]
\centering\includegraphics[width=\columnwidth]{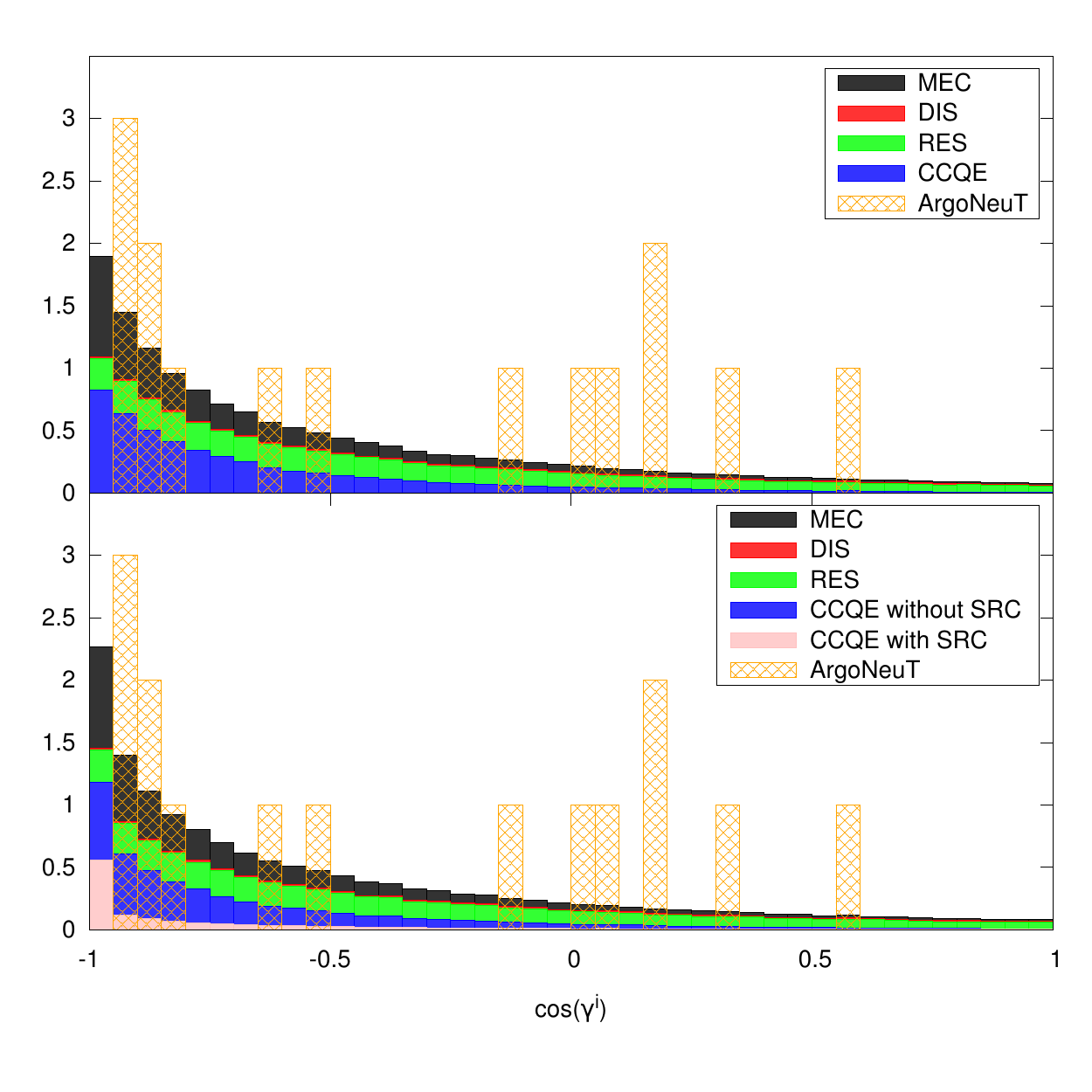}
\caption{(Color online) Distribution of the cosine of the reconstructed angle between two protons in the final state. 
(Top) LFG model, (bottom) SP approach.}
\label{fig:cos_g'}
\end{figure}

In Fig.~\ref{fig:cos_g'}, we compare the ArgoNeuT and NuWro distributions of $\cos\gamma^i$ normalized to the same area.

It is interesting to see that the NuWro distributions show a vast majority of events being reconstructed 
in the \mbox{back-to-back} initial state configuration. 
There are many reconstructed MEC events and also many SRC CCQE events in the NuWro SF mode. 
The appearance of many \mbox{back-to-back} SRC CCQE events in the SF mode is understandable, because in NuWro it is assumed 
that three-momentum transfer is absorbed by only one nucleon. The excess of MEC events may be surprising, because in NuWro 
the momentum transfered to the hadronic system is shared among both nucleons.

Using NuWro results as the probability distribution we calculate probabilities to have three (or more) events with 
$\cos\gamma^i\leq -0.9$ and six (or more) events with $\cos\gamma^i\leq -0.8$. 
The results for the two NuWro modes are shown in Table~\ref{tab:probabilities}.

\begin{table}[!htb]
\resizebox{.45\textwidth}{!}{
\begin{tabular}{l|c|c|}
\cline{2-3}
\rule{0pt}{2.2ex} & $\cos\gamma^i\leq -0.9$ & $\cos\gamma^i\leq -0.8$ \\ \hline
\multicolumn{1}{|l|}{NuWro: LFG} & \rule{0pt}{2.2ex} $P(3+)=65.0\%$ & $P(6+)=46.5\%$ \\ \hline
\multicolumn{1}{|l|}{NuWro: SF} & \rule{0pt}{2.2ex} $P(3+)=70.9\%$ & $P(6+)=50.6\%$ \\ \hline
\end{tabular}
}
\caption{Probabilities of detecting three (or more) and six (or more) events with protons 
in the reconstructed initial \mbox{back-to-back} configuration according to NuWro.}
\label{tab:probabilities}
\end{table}

We see that the enhancements in the \mbox{back-to-back} configurations of reconstructed nucleons is fully understandable in terms 
of NuWro simulations. 
The NuWro SF approach agrees with the data slightly better.
We notice also that the reconstructed \mbox{back-to-back} sample of events contains also a significant contribution from CCQE and RES events 
with no nucleon-nucleon initial state and two proton final state being the result of FSI effects. 
This suggests that there may be a general physical argument explaining the shape of the $\cos\gamma^i$ distribution. 
We will discuss this problem in Section \ref{sec:discussion}.

\section{Discussion}
\label{sec:discussion}

In the previous section our strategy was always to follow closely the ArgoNeuT procedures. 
Two main ArgoNeuT results were studied. Having at our disposal Monte Carlo generated events, 
we were able to discuss interaction modes contributing to experimentally selected samples of events~\cite{Acciarri:2014gev}. 
In this way we confirmed that hammer events in the LAB frame originate mostly 
from RES events, and that a substantial fraction of reconstructed \mbox{back-to-back} nucleon-nucleon pairs comes 
from the MEC mechanism and also from the CCQE mechanism on correlated nucleon-nucleon pairs.

In this section we would like to go deeper in the data/MC comparison study.

\subsection{Missing transverse momentum}
\label{sec:misstransmom}

The hammer events ($\cos(\gamma) \leq -0.95$) studied by the ArgoNeuT can be additionally characterized by the following 
conditions \footnote{Private communication from Ornella Palamara}:
\begin{itemize}
\item $|\vec{p}_1|, |\vec{p}_2| > k_F$,
\item $|\vec{p}^T_{miss}| \geq 220$~MeV/c,
\item $\frac{|\vec{p}_1|}{|\vec{p}_2|} \leq 1.2$.
\end{itemize}
In the case of the ArgoNeuT two proton sample, there are seven events (including four hammers) 
satisfying the above criteria ($7/30\approx 23\%$). In the case of
NuWro events, the subsample of only about $9\%$ of two proton events is accepted. 

\begin{figure}[!htb]
\centering
\includegraphics[width=\columnwidth]{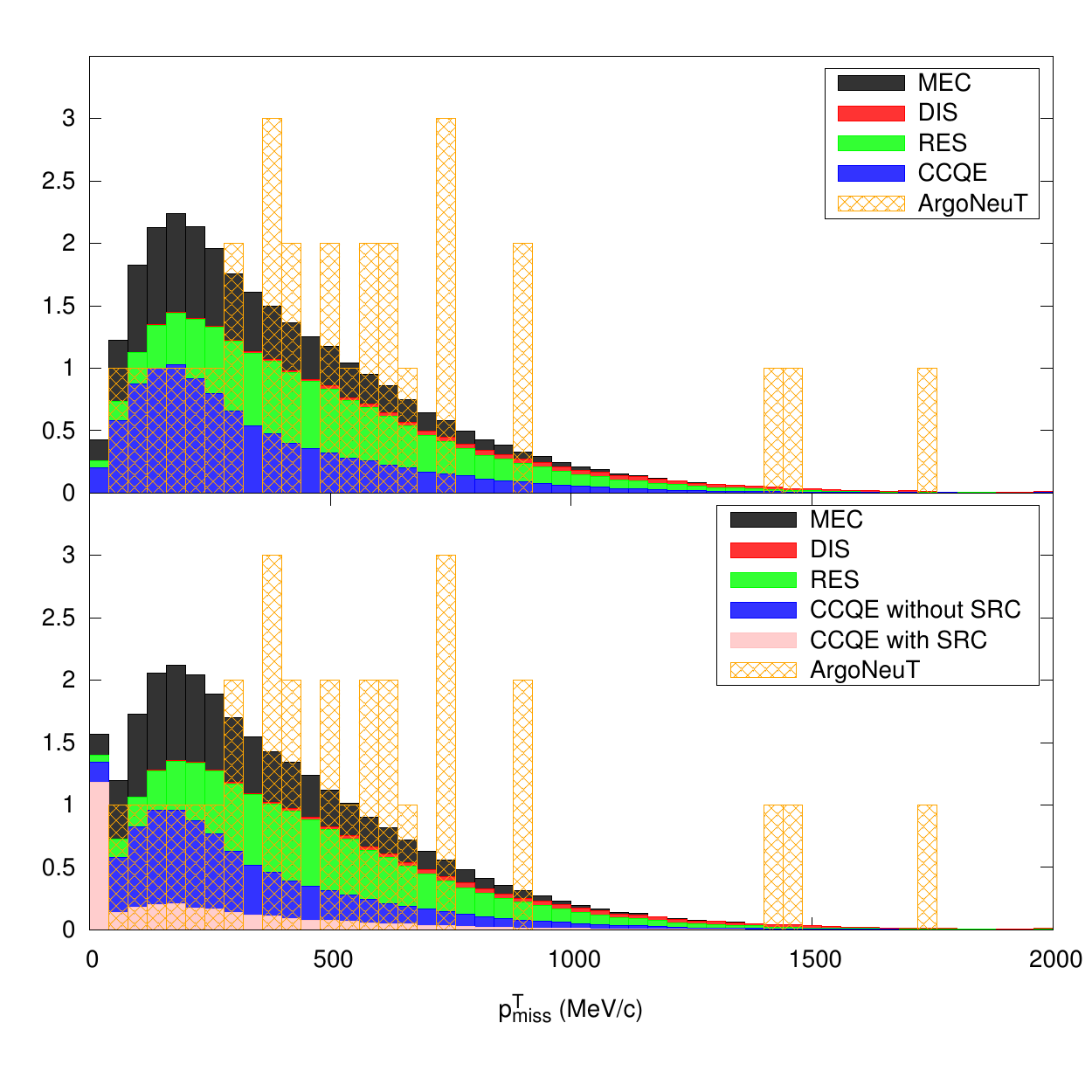}
\caption{(Color online) Missing transverse momentum distribution for 2 proton events. (Top) LFG model, (bottom) SP approach.}
\label{fig:missing_transverse_momentum}
\end{figure}

One of the additional conditions was defined in terms of missing transverse momentum. We investigated how well 
the measured distribution of missing transverse momentum is reproduced by NuWro. Fig.~\ref{fig:missing_transverse_momentum} 
shows a distribution of $p^T_{miss}$ from 29 ArgoNeuT two proton events (one of the events exceeded the histogram range). 
The distribution obtained with NuWro is shown also, and the NuWro results are normalized to the same histogram area. 
We notice the enhancement of the events in the first bin (almost zero $p^T_{miss}$) in the SF NuWro configuration. 
If both nucleons from the initial correlated nucleon pair did not suffer from FSI effects the missing transverse momentum 
is exactly zero (in the NuWro SF mode nucleons in the initial correlated state
have opposite three-momenta). We checked that large missing transfer momentum events contain high momentum neutrons. 
Fig.~\ref{fig:missing_transverse_momentum} suggests that NuWro understimates a probability to have such events.

\begin{figure}[!htb]
\centering
\includegraphics[width=\columnwidth]{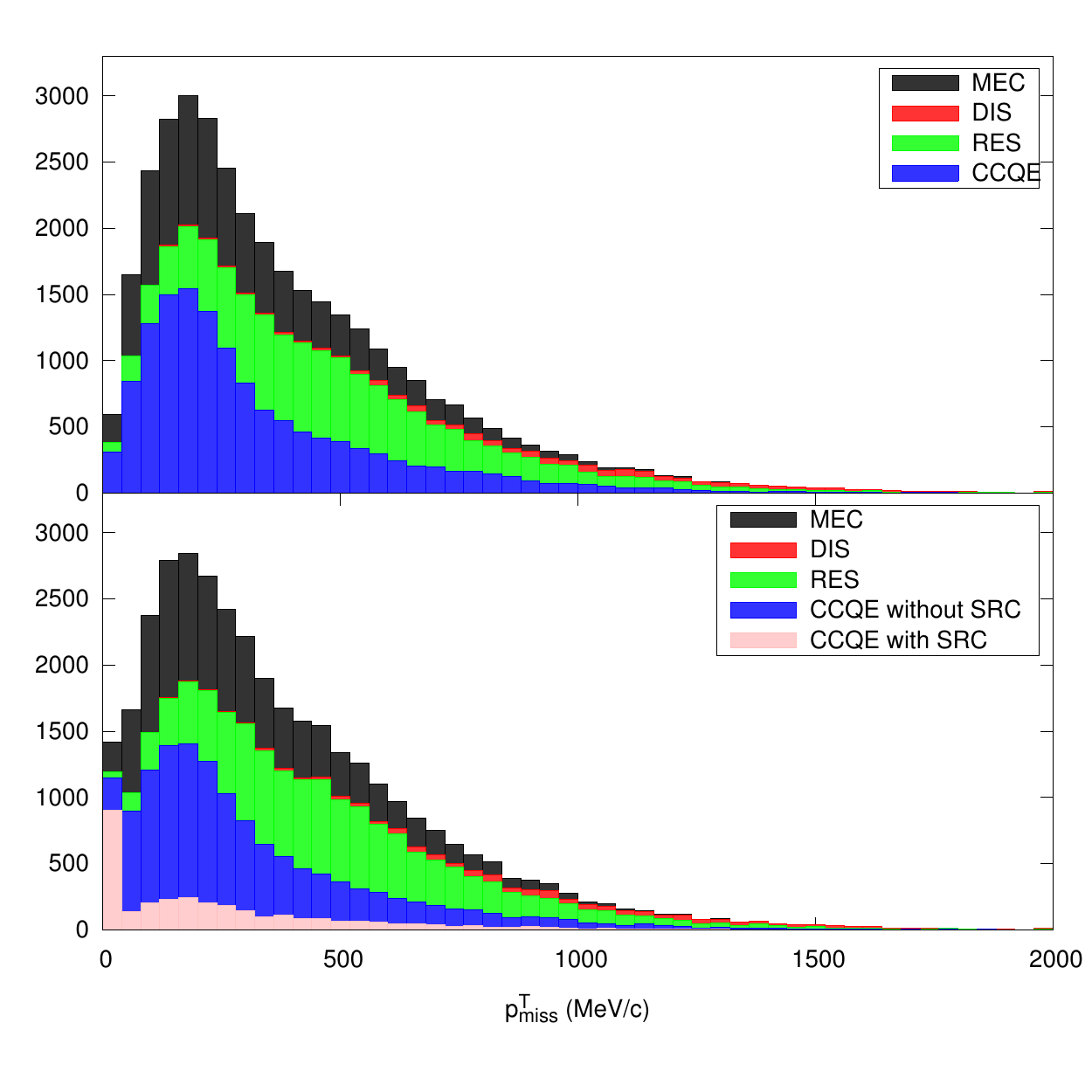}
\caption{(Color online) Missing transverse momentum distribution for 2 proton hammer events. (Top) LFG model, (bottom) SP approach.}
\label{fig:missing_transverse_momentum_hammer}
\end{figure}

We looked also at the NuWro distribution of $p^T_{miss}$ for the hammer events satisfying also 
$|\vec{p}_1|, |\vec{p}_2| > k_F$ and $\frac{|\vec{p}_1|}{|\vec{p}_2|} \leq 1.2$.
The distribution is shown in Fig~\ref{fig:missing_transverse_momentum_hammer}. 
It is interesting to see that for $p^T_{miss}\geq 300$~MeV/c,
the RES contribution starts to dominate. One should also expect many hammer events from CCQE and MEC mechanisms characterized
by $p^T_{miss}\sim 200$~MeV/c, however they are missing in the data. Better statistics experimental distribution of 
hammer events $p^T_{miss}$ could provide very useful information about CCQE/MEC and RES mechanisms separately.

\subsection{Kinematics reconstruction procedure}
\label{sec:kinrecpro}

We investigated 
how reliable the procedure is to reconstruct the initial nucleon-nucleon configuration in the restricted subsample of 15 events. 
We calculated distributions of differences between reconstructed and true values of three quantities: 
neutrino energy, the value of three-momentum transfer and the three-momentum transfer direction. 
The results are shown in Table~\ref{tab:reconstruction}, where $\mu$ is the mean value and $\sigma$ is the standard deviation.

\begin{table}[!htb]
\resizebox{.45\textwidth}{!}{
\begin{tabular}{l|c|c|}
\cline{2-3}
\rule{0pt}{2.2ex} & NuWro: LFG & NuWro: SF \\ \hline
\multicolumn{1}{|l|}{$\mu(E_\nu^\prime - E_\nu) \ [{\rm MeV}]$} & \rule{0pt}{2.2ex} $-241$ & $-238$ \\ \hline
\multicolumn{1}{|l|}{$\sigma(E_\nu^\prime - E_\nu) \ [{\rm MeV}]$} & \rule{0pt}{2.2ex} $488$ & $486$ \\ \hline
\multicolumn{1}{l|}{} & & \\ \hline
\multicolumn{1}{|l|}{$\mu(1 - \cos(\vec{q}^\prime, \vec{q}))$} & \rule{0pt}{2.2ex} $-0.04$ & $-0.04$ \\ \hline
\multicolumn{1}{|l|}{$\sigma(1 - \cos(\vec{q}^\prime, \vec{q}))$} & \rule{0pt}{2.2ex} $0.079$ & $0.079$ \\ \hline
\multicolumn{1}{l|}{} & & \\ \hline
\multicolumn{1}{|l|}{$\mu(|\vec{q}^\prime - \vec{q}|) \ [{\rm MeV/c}]$} & \rule{0pt}{2.2ex} $-244$ & $-242$ \\ \hline
\multicolumn{1}{|l|}{$\sigma(|\vec{q}^\prime - \vec{q}|) \ [{\rm MeV/c}]$} & \rule{0pt}{2.2ex} $488$ & $486$ \\ \hline
\end{tabular}
}
\caption{The biases and standard deviations of reconstruction of neutrino energy, three-momentum transfer direction and 
three-momentum transfer value. From Eq.~\ref{eq:reconstruction}, 
it is clear that the quality of reconstruction of neutrino energy and momentum transfer
are strongly correlated.}
\label{tab:reconstruction}
\end{table}

The reconstruction formula tends to underestimate the neutrino energy and therefore, the value of three-momentum transfer. 
This is due to presence of undetected neutrons in the final state. In larger liquid argon detectors like MicroBooNE, the
kinetic energy carried away by neutrons may be partially seen via interaction with visible energy deposit making the reconstruction more
precise. On the other hand, the direction of the three-momentum transfer is reconstructed quite accurately.

Finally, we would like to address the problem of the shape of the $\cos\gamma^i$ distribution. Momentum conservation implies that there is a 
correlation between $\vec q$, 
$\vec q_{rec}$ and both nucleon three-momenta: $\vec {p}_1$ and $\vec {p}_2$. 
Using NuWro, we checked that the distribution of $\cos (\vec q_{rec}, \vec p_1)$ is peaked at $\sim 0.85$. The distribution of 
$\cos (\vec q_{rec}, \vec p_2)$ is more diffused with a maximum at $\sim 0.6$.
The correlation between  $\vec q_{rec}$ and nucleon
three-momenta is smeared out by FSI effects and becomes weaker for events with larger number of nucleon rescatterings inside nucleus. 
Neglecting contribution from non-detected neutrons and nucleus recoil one can expect that $\vec q\approx \vec {p1} + \vec {p2}$ and 
$\vec q\approx \vec q_{rec}$. 
It is now clear that, if we define $\vec {p}^{\ i}_1\equiv \vec {p}_1 -\vec q_{rec}$, as done in the ArgoNeuT paper, 
we should expect that 
$\vec {p}^{\ i}_1$ and $\vec {p}_2$ will tend to be anti-parallel. Nuclear effects like Fermi motion and most importantly FSI
make the relation between $\vec {p}^{\ i}_1$ and $\vec {p}_2$ more complicated, but the basic feature of the distribution seen in 
Fig.~\ref{fig:cos_g'}, namely the peak in $\cos\gamma^i$ at -1, can be understood with presented above simple kinematical considerations.
The shape seen in Fig.~\ref{fig:cos_g'}
is universal and does not depend much on the dynamical mechanism behind the appearance of the two proton final state.

\section{Conclusions}
\label{sec:conclusions}

We followed the ArgNeuT study of the two proton and no pion events using simulations from the NuWro Monte Carlo event generator. 
NuWro has been used in the past in many experimental data studies, succesfully reproducing the measured results.

The most spectacular ArgoNeuT results is the appearance of several hammer-like events with almost \mbox{back-to-back} 
two proton configuration 
in the LAB frame. According to NuWro, the probability to have that many hammer events varies from $\sim 3\%$ to $\sim 1\%$, 
depending on how the observable is defined. These results 
suggest that an important physical mechanism leading to two proton 
and no pion final state may be missing in NuWro and quite likely also in other neutrino 
event generators. Better statistics data from awaited liquid argon MicroBooNE experiment will
allow us to understand better the situation. Useful information about physical processes can be obtained from the
missing transverse momentum distribution studies, allowing the examination of nuclear physics models implemented in MCs. 

Another interesting observable is reconstructed angle between two nucleon in the hypothetical initial nucleon-nucleon state. 
ArgoNeuT reported an access of \mbox{back-to-back} nucleons, and this fact can be understood using models implemented in NuWro.
We argued that the access is kinematical in origin and is not directly related with existence of SRC nucleon pairs.
Nevertheless, the details of the distribution shape is sensitive to SRC pairs, and it may be an important observable to
investigate in future experiments.

\begin{acknowledgments}
We thank Ornella Palamara and Flavio Cavanna for their interest in this study and for a lot of useful information about
experimental set up and comments.
JTS was partially supported by the NCN grant UMO-2014/14/M/ST2/00850. We thank M.A.R. Kaltenborn formproofreading the manuscript of the
paper.
\end{acknowledgments}

\bibliographystyle{apsrev4-1}

\bibliography{argoneut2}

\end{document}